Invited Paper

# Where, When, and How mmWave is Used in 5G and Beyond

Kei Sakaguchi[1, 2], *member*, Thomas Haustein[1], Sergio Barbarossa[3], Emilio Calvanese Strinati[4], Antonio Clemente[4], Giuseppe Destino[5], Aarno Pärssinen[5], Ilgyu Kim[6], Heesang Chung[6], Junhyeong Kim[6], Wilhelm Keusgen[1], Richard J. Weiler[1], Koji Takinami[7], *member*, Elena Ceci[3], Ali Sadri[8], Liang Xian[8], Alexander Maltsev[8], Gia Khanh Tran[2], *member*, Hiroaki Ogawa[2], *student-member*, Kim Mahler[1], Robert W. Heath Jr.[9], *member*

**SUMMARY**

Wireless engineers and business planners commonly raise the question on where, when, and how millimeter-wave (mmWave) will be used in 5G and beyond. Since the next generation network is not just a new radio access standard, but instead an integration of networks for vertical markets with diverse applications, answers to the question depend on scenarios and use cases to be deployed. This paper gives four 5G mmWave deployment examples and describes in chronological order the scenarios and use cases of their probable deployment, including expected system architectures and hardware prototypes. The paper starts with 28 GHz outdoor backhauling for fixed wireless access and moving hotspots, which will be demonstrated at the PyeongChang winter Olympic games in 2018. The second deployment example is a 60 GHz unlicensed indoor access system at the Tokyo-Narita airport, which is combined with Mobile Edge Computing (MEC) to enable ultra-high speed content download with low latency. The third example is mmWave mesh network to be used as a micro Radio Access Network (μ-RAN), for cost-effective backhauling of small-cell Base Stations (BSs) in dense urban scenarios. The last example is mmWave based Vehicular-to-Vehicular (V2V) and Vehicular-to-Everything (V2X) communications system, which enables automated driving by exchanging High Definition (HD) dynamic map information between cars and Roadside Units (RSUs). For 5G and beyond, mmWave and MEC will play important roles for a diverse set of applications that require both ultra-high data rate and low latency communications.

## 1. Introduction

5G will not be just a new radio access standard. Rather, the key novelty of 5G will be the integration of multiple networks serving diverse sectors, domains and applications, such as multimedia, virtual reality (VR) / augmented reality (AR), Machine-to-Machine (M2M) / Internet of Things (IoT), automotive, Smart City etc. [1][2]. A recent report states that data traffic for these new applications is expected to grow from 2016 to 2021 much larger than the assumed 590% for conventional applications: 770% for mobile video, 950% for mobile VR and 1320% for M2M/IoT [3]. The diversity of the 5G applications and the diversity of the related service requirements in terms of data rate, latency, reliability, and other parameters lead to the necessity for operators to provide a diverse set of 5G networks.

Among the key physical layer technologies enabling this efficient 5G system design, the use of mmWave spectrum will be coupled with network densification and massive MIMO to serve as an ultra-high speed access and backhaul systems. In this paper, we'll focus on the 28 GHz frequency band, which is planned to be used in 3GPP NR [4][5], and the 60 GHz band, which is an unlicensed band currently used by IEEE802.11ad/WiGig [6]. The rationale of this paper, though, can also be applied to future frequency bands being discussed in WRC-19 and later [7]. Another important 5G key technology is Mobile-edge Computing (MEC), which will bring information and processing closer to the mobile users and enable low latency services.

This paper provides an introduction to mmWave and MEC and introduces four different 5G mmW use cases, including a future forecast about where, when, and how mmWave will be used in 5G and beyond. The question on "how" was partly already answered: mmWave will be combined with MEC in order to allow ultra-high speed and low latency communications. The answer to the question on "where" will be disclosed in later sections with corresponding scenarios and use cases.

The paper is organized as follows. Section 2 summarizes related works and helps the reader to understand state-of-the-art of mmWave technologies. In Section 3 to 6, four different 5G mmW use cases are introduced, together with the expected time frame for market entry. Section 3 introduces 28 GHz outdoor backhaul for fixed wireless access and moving hotspot. Section 3 also introduces several Proof-of-Concept (PoC) demonstrations, which will be presented at the PyeongChang winter Olympics games in 2018. Section 4 introduces the trial demonstration of a 60 GHz indoor access system combined with MEC at the Tokyo-Narita airport. Section 5 introduces a mmWave mesh network for μ-RAN, to be used in dense urban scenarios as a cost-effective integration of access and backhaul. Section 6 introduces mmWave based V2V/V2X communications for automated driving, which is expected to be one of the largest applications of 5G and beyond. Finally, Section 7 provides concluding remarks of this paper.

---

[1]Fraunhofer HHI, Germany. [2]Tokyo Institute of Technology, Japan. [3]Sapienza University of Rome, Italy. [4]CEA/LETI Labs – MINATEC, France. [5]University of Oulu, Finland. [6]ETRI, Korea. [7]Panasonic Corporation, Japan. [8]Intel Corporation, USA. [9]The University of Texas at Austin.

## 2. Related Works on mmWave Technologies for 5G & Beyond

This section starts with the current status of mmWave technologies in cellular networks and also explains issues related to spectrum regulation. To the author's knowledge, the first studies considering mmWave as a key component for cellular 5G networks appeared around 2011. Samsung was an early pioneer in recognizing the feasibility of mmWave for access in cellular systems [8]. Prof. Rappaport and his team at NYU showed the feasibility of mmWave for outdoor access scenarios based on propagation measurements at 28, 38, 60 and 73 GHz [9]. Prof. Heath and his team at University of Texas at Austin showed the effectiveness of beamforming and MU-MIMO in mmWave bands for system rate improvements in cellular networks [10]. Prof. Sakaguchi, Dr. Haustein, and the MiWEBA project team advanced the concept of heterogeneous networks overlaying mmWave small-cells on larger macro-cells [11], which is one of the baseline system architecture of current 5G standardizations.

It is very likely that 28 GHz will be used for the first 5G deployments in the South Korea, US and Japan, although 28 GHz itself was excluded from the candidates of IMT bands in WRC-15. ITU-R selected at the WRC-15 several frequency bands above 24.25 GHz as candidates for 5G [12] based on the IMT feasibility study for bands above 6 GHz [13] and the current usage of spectrum in all nations. Then, the FCC in the U.S. opened in total nearly 11 GHz of spectrum above 27.5 GHz, including unlicensed spectrum at 64-71 GHz, which resulted in the US leadership of world-wide 5G developments [14][15]. The same year, 5GMF [16] and EC [17] revealed their plans for 5G frequency bands in Japan and Europe, which are summarized in Table 4 in Section 6. Even though the bands around 28 GHz are prioritized for 5G in the South Korea, US and Japan, there are obviously frequency band discrepancies in different nations. In order to solve this problem, discussion in 3GPP proposes to treat spectrum from 24.25-29.5 GHz as one single band and adapt the actually used channel based on the local regulation by numerology.

Due to its technology readiness level, 60 GHz WiGig is also an important technology for 5G systems. After the standardization of IEEE802.11ad/WiGig in 2012, tri-band chipsets supporting 2.4 GHz, 5 GHz and 60 GHz bands were released by Intel and Qualcomm [18][19]. Panasonic developed in 2014 a first WLAN access point (AP) prototype using 60 GHz [20] and recently, commercial products using WiGig became available on the market. To ensure interoperability, WiGig became part of the Wi-Fi alliance in 2013 and the certification program was established in 2016 [21]. As a result, one can expect that the first WiGig embedded smartphones will commercially available in 2017.

The 60 GHz band in combination with highly directional antennas is also beneficial for short range small-cells backhauling. For instance, Intel has developed a 25 dBi beam steerable antenna in [22], in order to extend the coverage of WiGig up to several hundred meters. InterDigital has released a product using 60 GHz for meshed backhaul with a maximum distance of 300 meters per link in [23], and has been extended by introducing Software Defined Network (SDN) to create a flexible path on the mesh network. Overall, mmWave backhauling /fronthauling is reasonable to support higher densification of small-cells and a flexible network, without the extensive costs of new cabling [24][25].

From chipset and antenna perspective, 28 GHz followed the 60 GHz band. Samsung led developments in the 28 GHz band together with NYU and developed early prototypes [26]. After the WRC-15, many chipset vendors started to develop 28 GHz technologies, which also led to the announcement of Intel to support 28 GHz in their 5G chipsets [27]. Qualcomm announced that the first commercial products featuring 5G NR modems expected to be available in 2019 and it will initially support the 28 GHz band with 800 MHz bandwidth via 8x100 MHz carrier aggregation [28]. IBM released 32-element dual polarized phased array with fully integrated transceiver IC in 28 GHz band [29], and NEC released a massive MIMO antenna at 28 GHz band with 500 elements, which will be used in the earliest deployment of 5G in Japan [30].

Recently, several organizations realized mmWave technology components and mmWave network integrations. The MiWEBA project has demonstrated the integration of LTE and a 60 GHz WiGig access system, using the LTE & WLAN aggregation protocol defined in 3GPP Rel. 13 [31]. The Tokyo-Narita international airport announced the launch of a Kiosk service in 2018 using 60 GHz WiGig technology, in order to download (or upload) large amount of data instantaneously [32]. In the US, Verizon and Samsung announced their service at 28 GHz, starting from September 2017 [33] and, together with Ericsson, Nokia, Cisco, Qualcomm and Samsung, the Verizon Technology Forum was initiated. T-Mobile and AT&T are performing similar trials in the US together with Ericsson, Nokia and Samsung. In South Korea, SK Telecom, Ericsson and BMW executed a trial of the 1$^{st}$ 5G connected car using 5G V2X [34], where they used the 28 GHz band to deliver UHD 4K video from a 360°-camera. In Japan, KDDI and Samsung performed handover experiments in the 28 GHz band [35]. Also, NTT docomo, together with Ericsson and Intel, are planning 5G trial environments using 28 GHz in Tokyo [36]. ETRI as a pioneer in millimeter-wave-based railway communications technology successfully gave the second field trial of Mobile Hotspot Network (MHN) prototype system at the Seoul subway line 8 in February 2017 [37], which was the world first millimeter-wave prototype system demonstrated in a running subway train with a peak data rate of 1.25 Gbps.

Although mmWave technologies at 28 GHz and 60 GHz have the potential to be used for early deployment of 5G in the South Korea, US and Japan, there is a need for deeper discussions regarding the scenarios, use cases and corresponding enabling technologies. We will start in Section 3 with the related 5G CHAMPION project [38], which will use 28 GHz technology at the PyeongChang winter Olympics games in 2018 [39] and we will then go to 60 GHz technology, which will be used at the Tokyo-Narita airport during the Tokyo Summer Olympics games in 2020 [40]. From Section 4, we will widen our scope based on the 5G-MiEdge project [41] and study the union of mmWave and MEC enabling ultra-high data rate and low latency applications. This concept is extended to future 5G topics, in Section 5 to a mmWave mesh network and in Section 6 to a mmWave based V2V/V2X system.

## 3. 28 GHz Outdoor Backhaul for Fixed Wireless Access and Moving Hotspots

3.1 Introduction to 28GHz Band

Today is not clear yet which mmWave bands will be first adopted by 5G technologies. Nevertheless, the ITU-R and 3GPP have aligned on a plan for two phases in the 5G standardization. The first phase will end in September 2018 and South Korea, US, and Japan have agreed to trial solutions on 28 GHz band having the challenging target to first roll out of 5G services in real environment at the 2018 PyeongChang winter Olympic games. This might push 28 GHz into consumer products before the standardization bodies finalize the 5G standards even if ITU-R excluded the 28 GHz band from the candidates for WRC-19 because such band is adopted today for satellite communications.

The choice of the 28 GHz band is motivated by several reasons. First, there is extensive licensed but underutilized mmWave spectrum around 28 GHz bands that have been shown to support cellular communications in the range of 500 meters [26]. Second, 28 GHz band is still useful to create multipath environments compared to higher frequencies and can be used for non-line-of-sight communications. Moreover, an additional important advantage of exploiting 28 GHz band for wireless backhauling is the possibility of reuse 3GPP LTE functionalities. For instance, 3GPP LTE allows to reuse the LTE physical layer, originally designed to operate at carrier frequencies around 2 GHz, and applied it to higher frequencies up to 40 GHz for small cell backhauling [5]. This requires small modification of the numerology to increase the subcarrier spacing. This result in a cost effective adaptation of existing technology to accommodate the new 5G requirements as well as the opportunity of a quick launch of new features for wireless backhaul links.

Recently, South Korea, US, and Japan have been speeding up development of 5G services that use 28 GHz under the umbrella of the 28 GHz initiative. It is already planned in South Korea to demonstrate world first 5G services in PyeongChang winter Olympic games in 2018. However, real deployment strategy of 28 GHz band for both static and mobile 5G service provisioning will be consolidated after the extensive real-field experimentation at the PyeongChang Olympic games.

5G CHAMPION project [38] takes the challenge of providing the first fully integrated and operational end-to-end 5G prototypes to the PyeongChang Olympic games in 2018. This effort is a major leap ahead compared to existing and planned punctual technology trials, such as PoC platforms focusing on 28 GHz mmWave communication. This section describes the overall set-up of the PoC in the 5G CHAPION project including a synergetic combination of technologies such as beamforming based mmWave, virtualized infrastructure, software reconfiguration across the entire stack, and high-speed solutions. More details on the foreseen PoC are given in [42].

3.2 Scenario/Use Cases and Requirements selected in 5G CHAMPION Project

The 5G CHAMPION project leverages cutting-edge solutions of mmWave backhauling, mmWave transceivers with reconfigurable antennas, localized evolved packet core supported by distributed or centralized MEC with caching, media and streaming functionalities into a unique platform capable to support a wide-range of 5G specific use cases, summarized in Table 1.

More specifically, the first use case refers to a stationary multi-RAT hotspot connected via mmWave backhaul to the 5G European testing network. The objective is to demonstrate the capability of providing broadband connectivity (100 Mbps minimum). In this regard, the proposed strategy is to utilize high capacity mmWave wireless backhaul to increase coverage and infrastructure density, heterogeneous radio access towards the end users, and distributed SDN and NFV in the virtual EPC to optimally orchestrate services.

The second use case tackles the feasibility of ultra-high data rate over mmWave link. In this case, the objective is to reach 20 Gbps data rate over mmWave link and, to this end, the challenge is to develop efficient hybrid analog-digital beamforming as well as higher order modulation solutions.

The third use case addresses the problem of content distribution in high-speed trains (up to 500 km/h) and real-time video streaming in slow/medium moving hotspots (bus, tram, etc.). To this end, key technologies are mmWave wireless backhaul, beam switching, MIMO and heterogeneous access, and distributed virtual EPC.

Finally, the last two use cases to be demonstrated in PyeongChang Olympic games focus on short-latency (<5 ms end-to-end latency, of which 2 ms latency is target over-the-air) and broadband application scenarios. In this regard, in addition to rate and delay requirements, we tackle another KPI, namely, 5G inter-systems interoperability, which is a pure core network KPI and consists of monitoring, managing and orchestrate different core networks implementations. In fact, the 5G CHAMPION project will showcase a unique 5G system PoC comprising of two interconnected 5G networks (one developed in South Korea and the other in Europe) operating in different frequency bands, using different air interface and core network specifications.

**Table 1** Use cases, target KPIs and 5G CHAMPION key technologies.

| Use case | Target test-bed | Target KPI |
|---|---|---|
| 1. Stationary multi-RAT hot-spot connected via mmWave backhaul to 5G European testing network | Stand-alone EU testbed, Oulu | 100 Mbps user exp. 2.5 Gbps mmWave backhaul Heterogeneous access SDN/NFV in EPC |
| 2. Ultra-high data rate over mmWave link | Stand-alone EU testbed, Oulu | 20 Gbps |
| 3. High user-mobility (simulation based) | Stand-alone, KR testbed | 2.5 Gbps mmWave wireless backhaul with 500 km/h mobility 100 Mbps user exp. |
| 4. Short-latency applications (e.g. multiplayer remote gaming and multi-remote control) | EU-KR 5G system, PyeongChang | Intersystem interoperability < 5 ms end-to-end latency 2.5 Gbps mmWave wireless backhaul SDN/NFV in virtual EPC |
| 5. Broadband applications (e.g., UHD video content delivery, 3D virtual reality) | EU-KR 5G system, PyeongChang | Intersystem interoperability (KR testbed) moving hot-spot (bus) 2.5 Gbps mmWave wireless backhaul SDN/NFV in virtual EPC |

### 3.3 Proof-of-Concept at PyeongChang Winter Olympic Games

3.3.1 Overall Concept of Proof-of-Concept

The 5G CHAMPION PoC will showcase 5G with 28 GHz dedicated technology during the PyeongChang winter Olympics games for use cases: (i) short latency applications, (ii) broadband applications with stationary, high and ultra-high user-mobility. Those two use cases permit to show how visitors at the Olympic games can experience 5G services in 5G relevant scenarios.

To this end, the European 5G network will be federated with the South Korean wireless backhaul testbed on-the-move that will be deployed at the IoT street in the Gangneung Coastal Cluster (GCC) near to the Olympics venues (see Figure 1) where some radio units will be deployed along the roadside where buses, provided with moving hot spots, pass by. People on the bus can experience the 5G services for the above mentioned use cases, i.e. short-latency and broadband with mobility. This trial requires intense development and testing of dedicated radio frequency frontends and antennas solutions to achieve (i) an agile interoperability of different 5G core network and radio access implementations; (ii) mmWave backhauling solutions for stationary and mobile scenarios; (iii) ubiquitous broadband/heterogeneous access; (iv) extreme real-time communications.

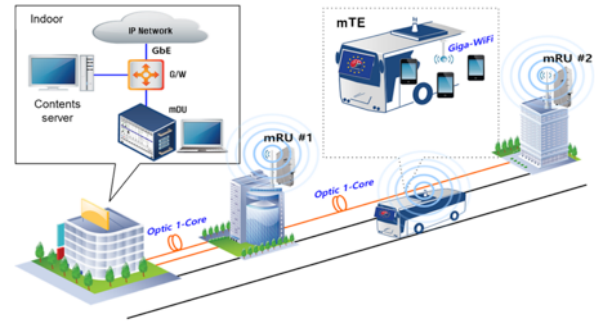

**Figure 1** PoC in PyeongChang Olympic games on backhaul to moving hotspots to be demonstrated at IoT street in South Korea.

3.3.2 28 GHz Technologies for Proof-of-Concept Demonstrations

Within the 5G CHAMPION project specific 28 GHz solutions have been selected for use in PoCs [44,45]. Two 28 GHz backhaul radio units will be developed, one in South Korea [43] and the other in Europe [45].

The basic system architecture of the South Korean backhaul solution for a MHN Enhancement (MHN-E) system entails two main components. The first, on the network side, is made of the MHN NodeBs (mNBs) consisting of MHN Digital Units (mDUs) and the MHN Radio Units (mRUs). The second, on the client- or the moving hot spot cell-side, is made of the MHN Terminal Equipment (mTE) mounted on a vehicle, which is primarily responsible for mobile wireless backhauling between the mRU in a mNB and the mTE. The mTE is connected to onboard access links (e.g., Wi-Fi) for the users inside a vehicle. The mRU is designed to operate in unlicensed frequency bands in the range of 25~26 GHz, which is referred to as the Flexible Access Common Spectrum (FACS) in South Korea.

In order to offer various broadband services inside a fast moving vehicle, the South Korean backhaul system adopts several advanced technologies [43], including a novel frame structure supporting carrier aggregation (CA) and enabling high-performance handover as well as digital MIMO technology using polarization antennas. As shown in [43], each aggregated carrier, generally referred to as a Component Carrier (CC), has a bandwidth of 125 MHz, and South Korean backhaul network allows the aggregation of a maximum of eight CCs to attain a total transmission bandwidth of up to 1 GHz. Besides, in order to further improve spectral efficiency, fixed-beam dual linearly-polarized antenna arrays with 4×4 (16 dBi gain) and 8×8 (21 dBi gain) elements are used on transmit and receive antennas respectively, to implement digital MIMO as shown in Figure 2.

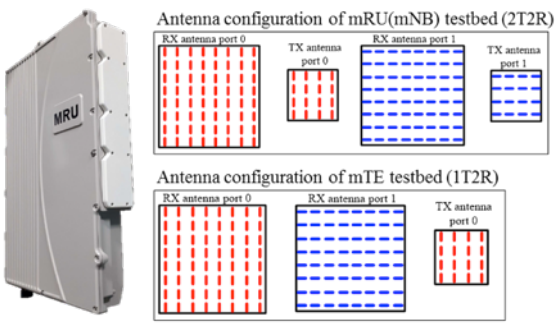

**Figure 2 Antenna configurations of South Korean 28 GHz backhaul testbed.**

In order to evaluate the link-level performance of South Korean backhaul network, a link-level simulation was conducted under the simulation parameters listed in the Table 2 and Cross Polarization Discrimination (XPD) of 50 dB referring to the recent ray-tracing simulation [44]..

**Table 2 Simulation assumptions for performance evaluation of South Korean backhaul network**

| Parameters | Setting |
| --- | --- |
| Channel model | Multipath-cluster channel model with Rician fading channel |
| K-factor | 20dB |
| Carrier frequency | 26GHz |
| Code rate | 0.8 |
| Transmission scheme | 2x2 MIMO |

The link-level simulation result in Figure 3 shows that the system is able to achieve average spectral efficiency of up to 10 bps/Hz in the case of 2x2 MIMO. Although only one mTE will be used during the demonstration in PyeongChang, it is still possible to achieve a data rate of up to 5 Gbps considering the system bandwidth of 1 GHz.

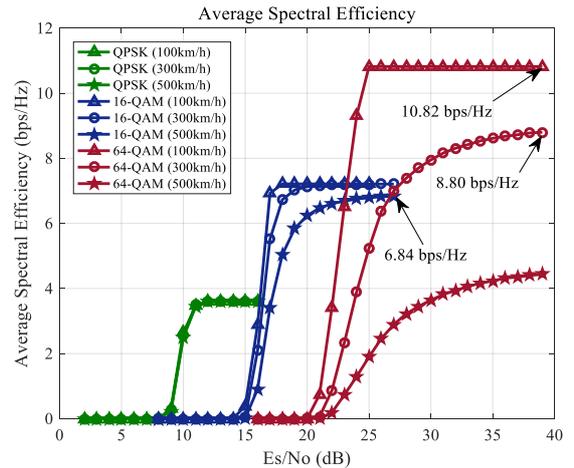

**Figure 3. Link-level performance evaluation of Korean backhaul network**

The European backhaul base station works in the frequency band 26.5 – 29.3 GHz and implements OFDM encoding and digital MIMO [44]. Analog beamforming will be implemented by considering two different electronically steerable antenna technologies, one based on the classical phased array architecture with phase shift inside the transceiver and the other considering a transmit array antenna working similar like a dielectric lens in optics but allowing adaptive beam forming electronically via a control interface.

The phased array solution is targeted for short to very long range operation with sophisticated power/dynamic range control both in TX and RX that would allow flexibility needed in backhauling to the local hotspot over a large area. RF hardware includes two separate antenna units each having dual-polarized antennas for four independent transceiver chains from the antenna into baseband. State-of-the-art power amplifiers can provide EIRP in the range of ~60 dBm or even beyond depending on the modulation for long range connections. The antenna gain of the phased array can be further increased using an external beam collimator that can be controlled electronically.

Figure 4 presents a block diagram of a possible solution for range boosting. More specifically, the antenna of backhaul base station can be designed with an electronically reconfigurable transmit array antenna with analogue beam-forming capability. A transmit array is typically composed of a focal source illuminating a flat-lens composed of phased unit-cells. By tuning the phase-shift on each unit-cell, for example integrating p-i-n diodes [45], the radiated beam can be collimated or deviated into a desired direction. The electronically steerable flat-lens is directly integrated into the radio front-end and feeding antennas are positioned in focal point of the electronic lens as in Figure 4.

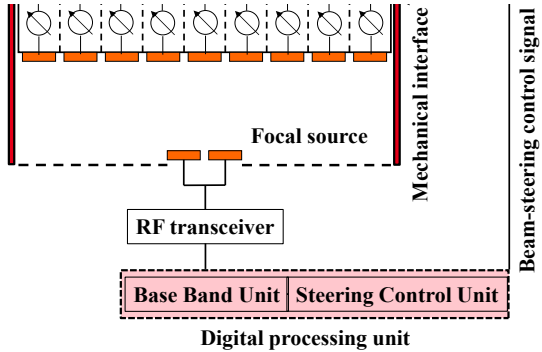

**Figure 4** Schematic view of the backhauling base station with electronically steerable transmit array

The analogue beamforming capability of a 400-element transmit array operating in the band 27.4 – 31.7 GHz has been recently demonstrated in [44]. The antenna is based on a 1-bit (two tunable phase-states) unit-cell [45] and works in circular polarization with a maximum broadside gain of 20.8 dBi. The measured scanning losses are equal to 2.5 and 5 dBi for a 40° and a 60° tilted beam, respectively. A linearly-polarized transmit array prototype working in the band 24 – 30 GHz is developing to cover both the South Korean and European frequency bands. The simulated results of its analogue beamforming capability are presented in Figure 5.

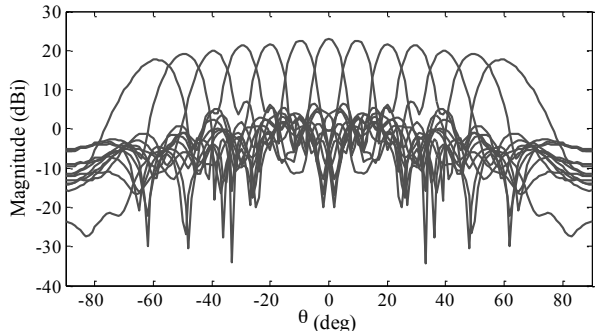

**Figure 5** Simulated analogue beamforming capability of the linearly-polarized transmit array developed in 5G CHAMPION at 28 GHz (E-plane gain radiation pattern).

## 4. 60 GHz Indoor Access with Mobile Edge Computing

4.1 Scenario/Use Cases and Requirements

mmWave access using 60 GHz unlicensed band in indoor private scenarios is anticipated to be deployed in near future. The scenarios do not only include homes and offices, but also areas such as airports, stations, trains, buses, stadiums, museum, and shopping malls, where mobile operators have difficulties to deploy their BSs directly. Although the size of such areas is limited, mobile users are expected to transfer large amount of video content at railway stations or airports, or use AR applications in stadiums and shopping malls. Therefore, in such an area, the combination of mmWave and MEC becomes very important to achieve both low latency and high data rate requirements. In this section, we list some use cases/scenarios where the combination of mmWave and MEC can enrich the user's quality of experience by following the discussions in the 5G-MiEdge project [41].

4.1.1 Cache Prefetching

To download large volumes of data using mmWave access, cache prefetching based on predictions of users' context is a key strategy to improve users experienced latency even with poor backhaul links. In some cases, caching may move together with the user, like a liquid. This may happen for example for users in a metro train downloading videos or large data files. In such a case, MEC servers associated to APs positioned in metro stations or along the metro rails may orchestrate video prefetching along the predicted positions of the users, so as to deliver these contents with very low latency. The information about radio channel capacity, number of connected users, and type of users' requests is fundamental to orchestrate the operation of nearby MEC assisted APs. Most of operations needed for services provisioning and network optimization are recurrent, context-dependent and too often re-executed. To exploit this property, [46] proposes an extension of the cache prefetching, the computational caching, in which power consumption and service delay of mobile edge computing are further reduced by caching popular computations.

4.1.2 Augmented Reality

AR services enrich a user's experience when entering a point of interest, like an airport, a museum or a sport event, by providing additional information to the user about what they are currently experiencing. AR applications need to be aware of the user's position and the direction they are facing through, for example, their camera view. Starting from such information, AR applications create additional information, in the form of video, sound, etc., and deliver it to the user in real-time. If the user moves, the information is refreshed and follows the user. This is a service that is naturally localized. It requires a high computational cost and low latency. Hosting such AR services on a MEC platform associated to a mmWave AP is a key strategy to create this computationally demanding supplementary information near the mobile user and deliver it within the required low latency requirements exploiting the very fast data transfer of mmWave links.

4.1.3 Computation offloading

Running computationally demanding applications on the mobile devices often leads to unpleasant user experience associated to the rapid discharge of the battery. Since the advancement of battery technology is not as fast as the development of new applications, a possible way out is to offload intensive computations on nearby servers, possibly located close to the access points. An effective deployment of computation offloading greatly benefits from the introduction of mmWave links and MEC servers able to orchestrate the deployment of virtual machines serving the users' needs when and where appropriate [47]. Computation offloading is also particularly useful to augment the capabilities of tiny sensors, in the IoT scenarios, which have very limited computation and storage resources.

4.2 Unification of mmWave Access and Mobile Edge Computing

MEC has been recently identified as an ecosystem enabling low latency and energy efficient proximity access to information technology services from mobile users [48]. The goal of MEC is to bring cloud-computing capabilities, including computing and caching, at the edge of the mobile network, within the Radio Access Network (RAN) and in close proximity to mobile users. This is obtained by empowering APs with additional storage and computation capabilities, and coordinating the work of nearby cloud-enhanced APs in order to serve the mobile users as efficiently as possible. This gives rise to a fully scalable application-centric system, where cache prefetching and computation offloading are brought as close as possible to the end user to reduce latency and energy consumption. Clearly, MEC is most effective in delivering context-aware services to mobile users, but it also helps deploying an effective caching and distributed computing strategy.

At the physical layer, some of the key technologies enabling the large data rate increase foreseen in the 5G roadmap are: dense deployment of radio APs, massive multi-input/multi-output (MIMO), and mmWave communications. Merging these physical layer technologies with MEC creates a very powerful system.

Figure 6 shows a system example where mobile users (equipped with low data rate 2.4/5 GHz WLAN or LTE) can get proximity access to information technology services, managed by MEC servers installed in nearby APs, through high data rate mmWave links. At the same time, mmWave links facilitate the orchestration of multiple MEC servers through high data rate mmWave backhaul connecting cloud-enhanced APs, possibly complementing wired backhauling [49]. Clearly, an effective cache prefetching or instantiation of virtual machines serving mobile applications requires local prediction of users' behavior and accurate estimation of popularity indices of most downloaded files, possibly varying over time [50]. This requires the extraction of big data analytics in cloud-enhanced APs, possibly coordinated through high data rate mmWave backhaul links.

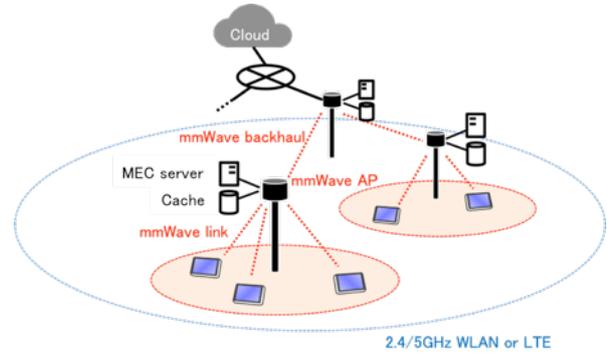

**Figure 6: System architecture example composed of mmWave access and mobile edge computing**

Adopting a user/application-centric point of view, if a user is accessing the network to run an application remotely, the most important thing is that the user gets a service with a good Quality of Experience, say for example low latency. Latency, as perceived by the user, includes time delay for data transfer plus the time necessary to run the application, plus maybe access to remote files. This suggests that the selection of the AP as well as the server where to instantiate the virtual machines running the application should be optimized *jointly* under a global latency constraint.

The use of multiple mmWave links clearly helps in providing fast access to local cloud computing services to multiple users distributed in a target area simultaneously. One of the impairments of mmWave links is channel intermittency, due to sporadic blockage due to obstacles or interference from mobile users observed from very similar angles by the APs. To counteract blocking events, it is useful to establish multilink communications, so that a user may access multiple APs at the same time, depending on channel conditions [20][51].

4.3 Prototype of mmWave Access with MEC at Narita Airport

As a preliminary work toward MEC through mmWave links, a multi-user mmWave access employing IEEE 802.11ad / WiGig [6] based APs is developed [52]. Figure 7 shows the AP prototype. The RF module integrates 60 GHz transceiver chip set with four element Tx/Rx beamforming, which provides about 120° beam steering range. Combining three RF modules, the AP achieves 360° wide area coverage while providing up to three concurrent links by using either the same frequency channel or different frequency channels. Three RF modules and the control board (which integrates a CPU, SDRAM, external Eathernet interface, etc.) are housed in the 80 x 150 x 150 mm unit. Each RF module can handle up to four stations (STAs), therefore accommodating a maximum of 12 users by the single AP. Figure 8 shows the measured output power of the RF module in the azimuth direction, offering 7-step beam steering with +20 dBm maximum effective isotropic radiated power (EIRP).

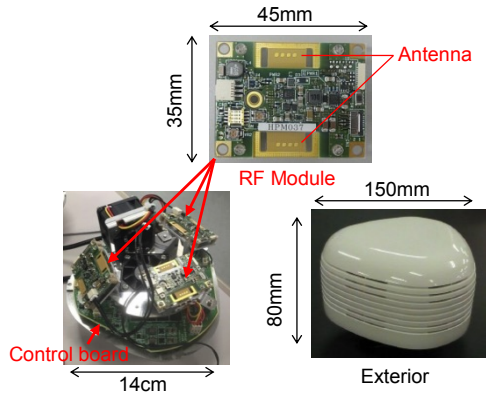

**Figure 7: Prototype of 60 GHz IEEE802.11ad / WiGig access point**

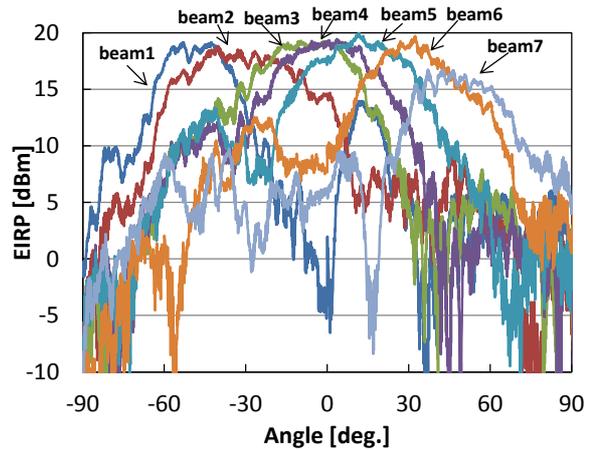

**Figure 8: Measured output power with different beams (at 58.32 GHz)**

Figure 9 shows the prototype network system. Three APs are wired with 10 Gbps Ethernet cables. The local content server acting as a edge cloud (cache) in this system bundles four 10 Gbps Ethernet cables to achive 40 Gb/s maximum throughput. The AP controller (APC) manages beamforming control and handover between RF modules and multiple APs. This is done by collecting Rx signal quality (signal-to-noise ratio) for each beam direction periodically, and select the opitmum mmWave link based on the beam routing table which includes the sector index (i.e. the beam direction index), MCS (modulation and coding scheme) and the Rx signal quality [20][52].

The prototype network system was set up in Narita International Airport as shown in Figure 10. Three APs were installed at a 2.5 m height, providing 10 x 5 m area coverage. To minimize interference among multiple APs and STAs, different frequency channels were allocated for each of the three RF modules within the APs. As STAs, nine 4K tablets were placed on counter desks at a 1.0 m height. They are equipped with IEEE 802.11ad / WiGig USB dongle prototypes to establish ultra-high speed 60 GHz wireless links with APs.

The STA achieves 1.7 Gbps maximum throughput, enabling to download a compressed two hours high-resolution video content (2 GB) within 10 seconds. During the nine day open period, 816 participants joined the experimental demonstration, and 99.3% of the positive feedbacks (high expectation for practical realization) were obtained. Future work includes architecture design of the whole system to the cloud, integration of caching/prefetching capability, etc. in order to realize ultra-high speed and low latency service delivery, which is resilient to network bottleneck such as backhaul congestion to the cloud.

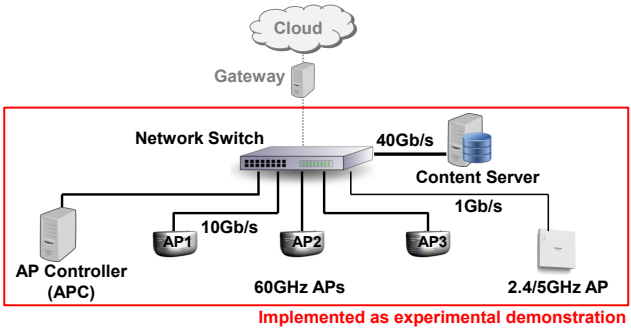

**Figure 9: Prototype network system with 60GHz access and edge content server**

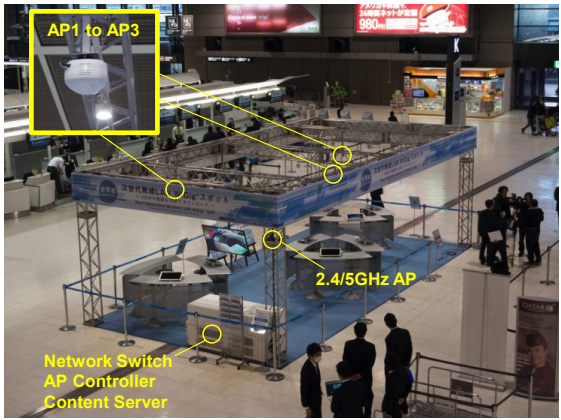

**Figure 10: Experimental Demonstration at Narita International Airport**

## 5. mmWave Mesh Networks for μ–RAN

### 5.1 Scenario/Use Cases and Requirements

In dense urban scenario which is one of the important scenarios in 5G, network densification is necessary because of the high traffic volume generated not only by smart phones and tablets but also by augmented reality information such as sensors and wirelessly connected cameras. Typical environments are open squares, street canyons, stations, etc., where users tend to gather and move as large and dynamic crowds while want to keep connectivity to the cloud. Figure 11 shows one example of such scenario around Shibuya station in metropolitan Tokyo.

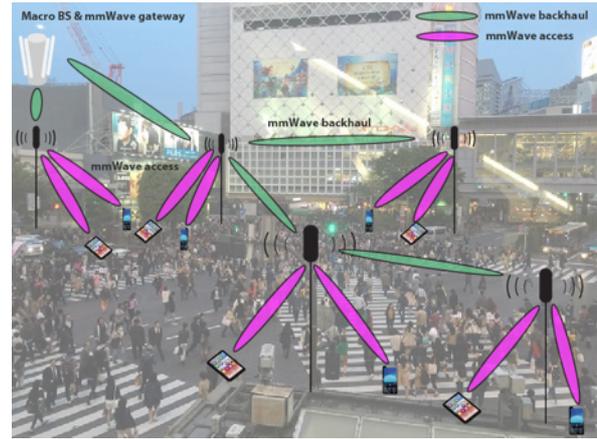

**Figure 11: An example of dense urban scenario around Shibuya station in metropolitan Tokyo**

Figure 12 and Figure 13 show measured data of mobile traffic in 2014 around a famous station in metropolitan Tokyo [53]. Figure 12 shows the spatial traffic distribution in one hour in that area from 10:00 to 10:59 AM. From the figure, it is obvious that the traffic distribution was not uniform and there were several hotspots. Figure 13 shows the time variation of the total traffic in this area. Since it is around an urban station, the traffic in midnight was very low, while that of daytime was very high. In this measurement, the average traffic demand per user was about 62 kbps and total area traffic at peak hour was 44 Mbps. In the following simulation examples, these numbers are multiplied by 1000 by considering future traffic in the era of 5G.

In such an environment, network densification with many number of mmWave small-cell BSs (SC-BSs) overlaid on the current LTE cells is effective to accommodate traffic in peak hours as drawn on Figure 11 [11]. However, many number of SC-BSs leads to the problem of high capital expenditures (CAPEX) and operating expenses (OPEX). One solution to relax the problem is to use mmWave mesh network for the backhaul network of SC-BSs as in the 5G-MiEdge project [41]. By using the mmWave mesh network, the CAPEX can be reduced by removing deployment cost of wired backhaul. Furthermore, the OPEX can also be reduced by introducing dynamic ON/OFF and flexible path creation in the backhaul network in accordance with the time variant and spatially non-uniform traffic distributions [54]. Such flexible control of the backhaul network is enabled by SDN technology using out-band control interface over the LTE [55][56]. In summary, mmWave mesh backhaul with SDN comes into place as one suitable candidate for dense urban scenarios owing to its ultra-wide bandwidth and deployment flexibility with low cost.

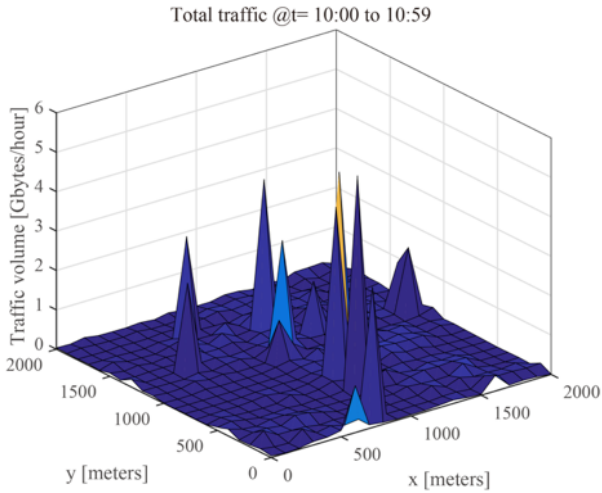

**Figure 12: Spatial distribution of the measured traffic in one hour**

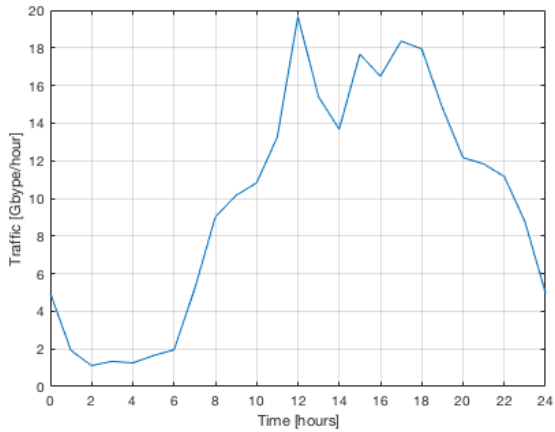

**Figure 13: Time variation of the measured traffic throughout a day**

5.2 mmWave Mesh Network with Traffic & Energy Management Algorithm

Figure 14 shows an example of mmWave mesh network to be used in the dense urban scenario. In Figure 14, mmWave SC-BSs are overlaid on a LTE macro cell to play a role of integrated backhaul and access with three or four sectors in both access and backhaul. The LTE macro BS plays a role of mmWave gateway as well in the cell to accommodate time-variant and spatially non-uniform traffic by forming a mmWave mesh network. Namely, a set of LTE macro BS and mmWave SC-BSs forms a μ-RAN for the target environment. The prominent objective of the traffic & energy management algorithm is to reduce energy consumption of mmWave mesh network by switching off as many SC-BSs as possible in an area while satisfying users' traffic demands. One example of such traffic & energy management algorithm proposed in [54] is shown in Figure 15. As it is hard to optimize ON/OFF status of SC-BSs and backhaul paths all at once, the algorithm involves three steps. In the first step (i), the initial ON/OFF status of SC-BSs is determined based on the traffic demands per SC-BS. In the next step (ii), initial paths of backhaul network are created to minimize power consumption. If isolated SC-BSs exist even after step (ii), the final step (iii) re-activates remaining SC-BSs in an energy efficient manner so as to transfer the traffic for the isolated SC-BSs. Control signaling to manage ON/OFF status of SC-BSs and to create physical paths between them are transmitted over the LTE as an out-band control plane. As such, a dynamic and energy efficient mmWave mesh network is formed.

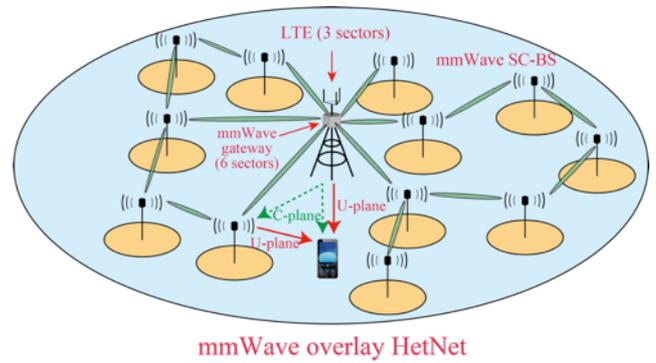

**Figure 14: mmWave mesh network overlaid on a macro cell**

(i) Initial ON/OFF status selection
**Goal:** Minimize the total power consumption of mmWave network
**Means:** In order to deactivate as many mmWave SC-BSs as possible, accommodate users in underloaded area to LTE within available bandwidth

(ii) Initial path creation for backhaul network
**Goal:** Satisfy user's traffic demand
**Means:** Distribute locally intensive traffic across multiple backhaul sources, and form backhaul path relaying only mmWave SC-BSs activated in (i)

(iii) Reactivation & path creation for isolated SC-BSs
**Goal:** Assure backhaul links for all mmWave SC-BSs
**Means:** Find the best combination of reactivated mmWave SC-BSs to relay for isolated SC-BSs so that the minimum number of SC-BSs are reactivated

**Figure 15: Traffic & energy management algorithm**

5.3 Simulation of mmWave Mesh Network

This section shows several examples of simulation analysis for mmWave mesh networks controlled by the abovementioned algorithm. In this numerical analysis, several macro cells with ISD of 500 m are assumed to be deployed within the 2000 m square areas in Figure 12 and one macro cell is selected as the evaluation cell. Other simulation parameters are shown in Table 3.

Table 3: Simulation parameters.

| Parameter | LTE | mmWave SC-BS |
|---|---|---|
| Bandwidth | 10 MHz | 2×2.16 GHz |
| Carrier freq. | 2.0 GHz | 60 GHz |
| Antenna gain | 17 dBi | 26 dBi |
| Antenna height | 25 m | 4m/25m (SC-BS/GW) |
| Tx power | 46 dBm | 10 dBm |
| Beam pattern | 3GPP | IEEE802.11ad |
| Path loss | 3GPP | [11] |
| # of BSs | 1 | 90 |
| Noise density | 174 dBm/Hz | |

Examples of the formed mmWave mesh networks are shown in Figure 16. As there are few users in the evaluation area at 3 AM, only a few mmWave SC-BSs are activated. In this case, since there are enough resource blocks in the LTE, most of the users are connected to the macro BS, while users with very high traffic demand at the right-bottom activate SC-BSs. On the other hand around 3 PM, a hotspot appears in the upper-left zone. We can see some backhaul links formed from gateway to the hotspot, showing the effectiveness of the traffic & energy management algorithm against the locally intensive traffic.

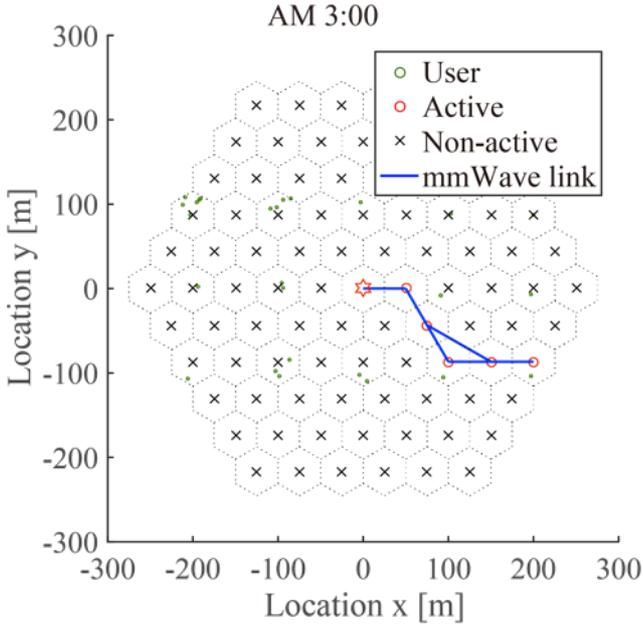

(a) Formed mmWave mesh network at AM 3:00.

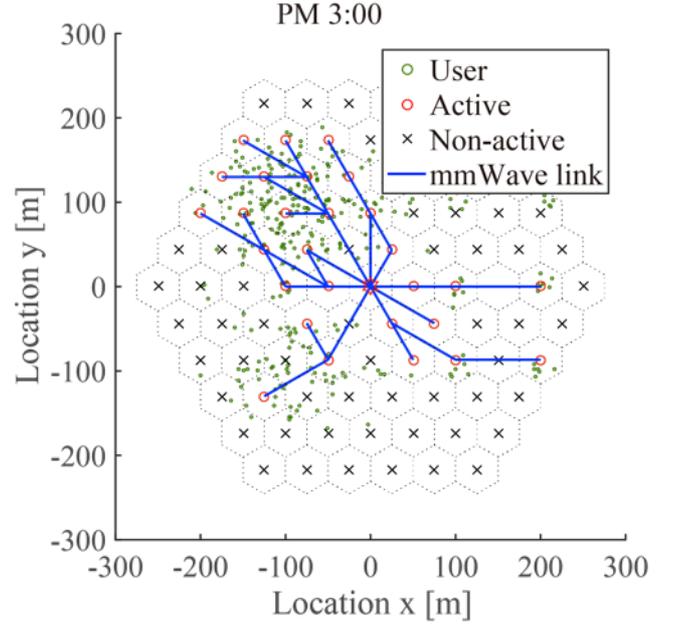

(b) Formed mmWave mesh network at PM 3:00.

**Figure 16: Dynamic formation of mmWave mesh network**

The analysis of power consumption is shown in Figure 17. Here, three types of criteria for SC-BS activation are compared. The first one is "Network centric ON" shown in Section 5.2. The second is "User centric ON" in which mmWave SC-BSs are turned on based on the location of users regardless of the traffic demand. The last one is "Always ON" without considering ON/OFF switching. The figure shows the performance of total power consumption against dynamic traffic variation throughout a day. The power consumption includes both of the access and backhaul, which is defined as follows,

$$\text{Energy consumption} = \sum_{i}^{N_{\text{AP}}} \left( N_i^{\text{on}} P_{\text{on}} + N_i^{\text{off}} P_{\text{off}} \right)$$

where $N_{\text{AP}}, N_i^{\text{on}}$, and $N_i^{\text{off}}$ represent the number of mmWave SC-BSs, the number of ON sectors and that of OFF sectors of $i$ th mmWave SC-BS respectively. From the figure, the effectiveness of the traffic & energy management algorithm is obvious especially in midnight. It is also noted that the Network centric ON reduces the energy consumption about a half by control the association point of users adaptively within the μ-RAN.

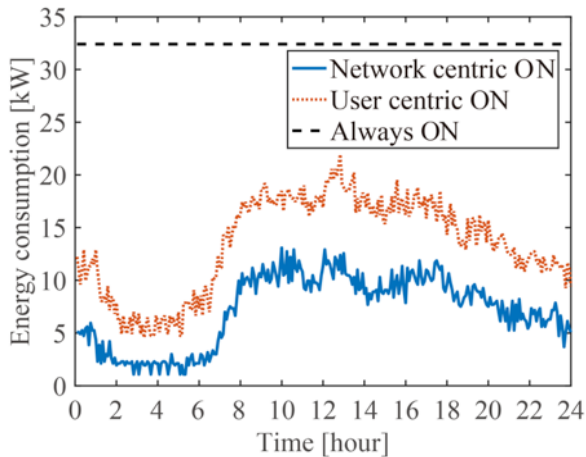

**Figure 17: Performance of total energy consumption**

5.4 Prototype of mmWave Mesh Nodes

At the last part of this section, we'll introduce prototype hardware to be used as the mmWave mesh node. Figure 18 is showing the latest compact universal mmWave platform supporting the IEEE 802.11ad/WiGig [6] developed by Intel. This mmWave PoC platform can be used both for flexible backhauling with point-to-point symmetrical transmission links up to 400 m coverage with capability of adaptive beam switching and also for access with asymmetrical transmission links between AP and STAs with limited capability of beam gain at the STA sides supporting coverage up to 100 m with the same throughput.

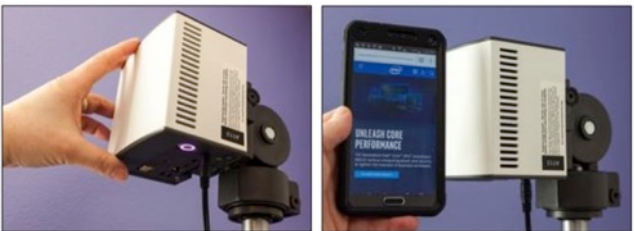

**Figure 18 Prototype hardware for mmWave mesh node**

Figure 19 shows the antenna geometry of the mmWave PoC platforms. This generation is designed with compact form factor, new radio modules, better antenna characteristics, and in-build mini PC (Intel i5 NUC) with Linux device drivers. The WiGig modem on the mini PC drives 8 radio modules with 16 antenna elements each. The 8 radio modules are used jointly to form a 128-element antenna array in a cost efficient way. A FPGA is used to process the commands from the WiGig modem, and send different commands to 8 radio modules for packet transmission, beamforming, etc. As a result, the prototype has about 41 dBm EIRP, 6 degrees azimuth beam width, 10 degrees elevation beam width, and +/-60 degrees azimuth and +/-30 degrees elevation steering ability. Some antenna steering patterns are shown in Figure 20. It consumes about 20 W system power. At 200 m in line-of-sight distance, up to 2 Gbps IP throughput could be achieved.

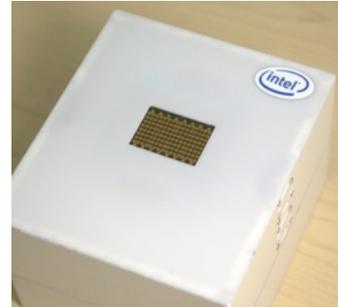

**Figure 19 Antenna geometry of mmWave PoC platform**

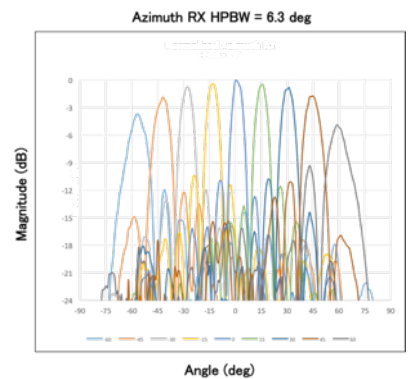

(a) Azimuth patterns.

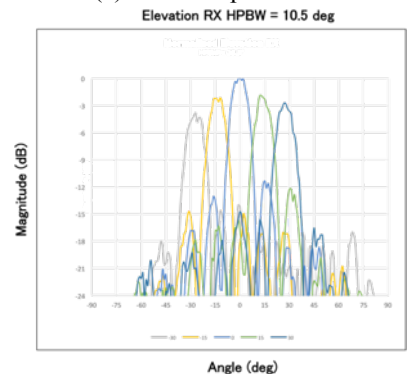

(b) Elevation patterns.

**Figure 20 Measured antenna steering patterns with different beams**

Finally, Figure 21 shows a photo of mini mmWave mesh network using six mmWave PoC platforms exhibited at Mobile World Congress 2017. At the center of the photo, there is a streetlight with Point-of-Presence (POP) of fiber backhaul. On the top of streetlight, three mmWave platforms are used to act as a gateway of mmWave networks by providing three mmWave backhaul links. The other three platforms are located at a SC-BS (right), at a home (left top), and at a drone (left) to be connected with the mmWave gateway via backhaul links. It is planned to integrate SDN capability on this platform in near future.

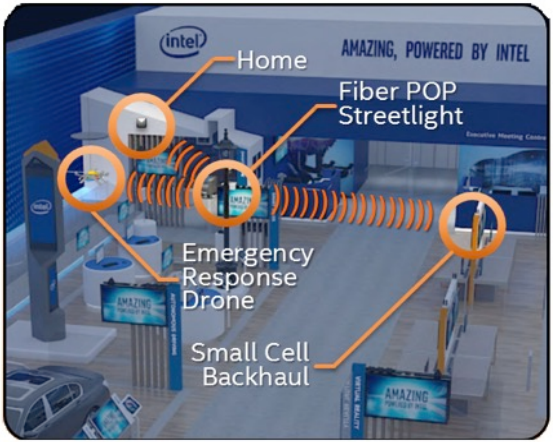

**Figure 21 mmWave mesh node to provide three different backhaul links exhibited in MWC2017**

## 6. mmWave based V2V/V2X for Automated Driving

6.1 Scenario/Use Cases and Requirements

Automated driving is one of the three most important applications of future 5G systems [57]. The $1^{st}$ phase of 5G Vehicle-to-Vehicle (V2V) and Vehicle-to-Everything (V2X) communications aims at driver assistance systems and exchanges messages either directly between vehicles or via appropriate infrastructure [58]. These messages are transmitted in case of an emergency or as so-called awareness messages, which contain information such as location, speed and heading direction. The $2^{nd}$ phase of 5G V2X aims for automated driving applications, where automated control software become primarily responsible for monitoring the environment and the driving vehicles, referred to as Levels of Automation (LoA) in the range 3 to 5 [59].

Automated driving systems require highly resolved and dynamic maps to maneuver the vehicles safely, in particular as a means to provide decimeter localization which is not achieved by typical consumer-grade satellite navigation equipment. As the resolution of current 2D maps coupled with inaccurate position information is not sufficient, high resolution and real-time maps, also called dynamic High Definition (HD) maps, become indispensable [60]. Figure 22 shows an example of an HD map generated with a LiDAR (Light Detection And Ranging) sensor, which is used to monitor the car surroundings and display the same as a high-resolution and real-time point cloud. It is reported in [60] that the total data volume of such an HD map collected for the duration of one hour is about 1 TB, which corresponds to a 2.2 Gbps data output for this type of sensor. It is reported in [61] that Google's automated car gathers a total of about 6 Gbps of sensor data, which includes LiDAR and other sensors, in order to out automated driving into practice. [62] summarizes in general for automated driving the required data rates of automobile sensors, i.e. 80-560 Mbps for LiDAR sensors or 160-320 Mbps for cameras sensors. Even though the final values depend on the resolution of the actual HD map, an estimation of 1 Gbps appears reasonable as a typical data rate requirement for exchanging HD map information via V2V/V2X links. Indeed, mmWave seems to be the only means to provide such high data rates to vehicles [71].

Cooperative perception is realized by exchanging sensor data between vehicles and Roadside Units (RSUs) and is necessary in order to widen or enhance the visibility area of HD maps [63][64]. This concept allows a more accurate localization of objects and more important, due to the bird's eye view that can be derived from sensor fusion, prevents that objects remain not visible and undetected, due to the ego-perspective of the sensors mounted on the vehicle. This external object detection capability is particularly critical for a safe realization of automated driving in complex urban environments. The communication range supported by a cooperative perception system can be determined by the braking distance of vehicles. As an example, [65][66] estimate 100 meters as a braking distance (including a safety increment distance) for the emergency stop of a generic sedan car at 70 km/h, which can be seen as the maximum speed in urban city environments. This braking distance estimate could be raised to 150 meters for buses or trucks. Hence, the HD map exchange system should support a communication range of at least 150 m for emergency braking applications. This number is to some extent comparable to the numbers given by [67].

At the end, latency is the most crucial communication system parameter in order to realize a stable control. It's well known that latency and data rate are a tradeoff for the case of video or LiDAR data transmission. The latest video data compression techniques reduce the data rate to one-tenth of the raw data rate. Video compression though leads to latencies higher than the 10 ms required for automated driving [57]. Furthermore, raw (or nearly raw) data is likely to be needed for interpretation by machine learning algorithms and has additional value for liability reasons in case of an accident [63].

In summary, the enhanced V2V/V2X communication targeting automated driving requires a data rate of 1 Gbps per link, end-to-end latency of less than 10 ms per link and a communication range of 150-300 m, to put safe automated driving into practice by exchanging raw (or lightly processed) sensor data. Such high requirements cannot be realized with current technologies [57]. Hence, 3GPP initiated related work in Release 15 and beyond, which is named eV2X (enhanced V2X) [68]. As a consequence, the utilization of mmWave and MEC technologies are becoming increasingly important for the field of automated driving.

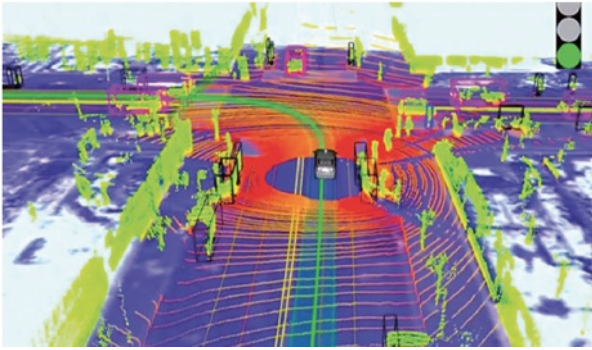

**Figure 22: HD map measured by LiDAR as a high-resolution point cloud [60]**

6.2 mmWave based V2V/V2X

Figure 23 shows an example of a system architecture for mmWave based V2V/V2X to realize a real-time exchange of HD maps between On-Board Unit (OBUs) mounted in vehicles and RSUs. All communication links between OBUs and RSUs are directly connected through Device-to-Device (D2D) communication, as specified in [69]. However, different from standard proximity-based services, this V2V/V2X system uses mmWave channels to fulfill the requirements of 1 Gbps data rate and less than 10 ms latency. The communication range of mmWave links can be extended to more than 300 m, if highly directional antenna beams are used. The beams can come from antennas with a small form factor, due to the small wavelength in mmWave bands [31]. One challenge in this system might be the antenna beam alignment between OBUs and RSUs, however [70][71] show the feasibility of using mmWave in vehicular scenarios. The system coexists with the conventional V2X system [58], which supports cloud-based services such as traffic jam forecast and long-range traffic navigation.

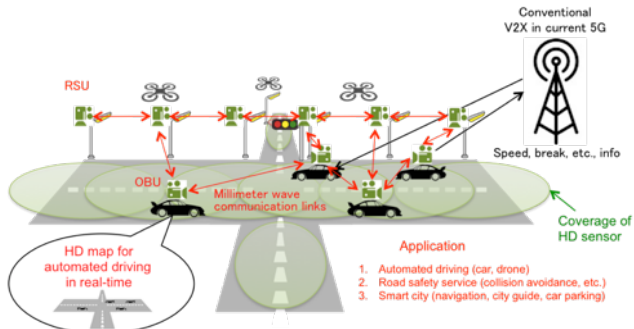

**Figure 23: mmWave based V2V/V2X to exchange HD maps**

A block diagram of OBU and RSU is shown in Figure 24, where the difference between a OBU and a RSU is solely the automated driving unit. The OBU/RSU receives via mmWave V2V/V2X links the HD maps from surrounding OBU/RSUs and fuses them with its own HD sensor data in the HD map processing unit. This process is called cooperative perception as described in Sect. 5.1. This combined HD map with its widened visibility area is used for automated driving decisions and in addition is transferred to neighboring OBU/RSUs. However, before transmitting the fused HD map, the HD map processing unit selects the area of interest (or control resolution of HD map area by area) dependent on the location of receiver OBU/RSUs to avoid exponential increase of data rate. The OBU/RSU is a unification of mmWave and MEC, since the HD map processing unit is considered as MEC to compute cooperative perception at the edge of the network.

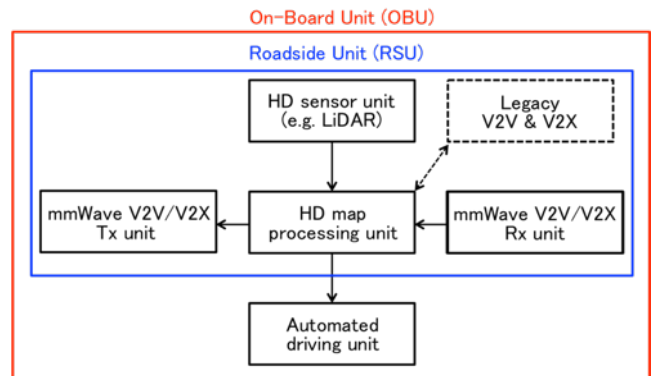

**Figure 24: RSU and OBU composed of mmWave and MEC**

Figure 25 shows an HD map example as a result of cooperative perception from multiple RSUs continuously monitoring the road conditions. In this case, the RSUs are located on the street lamps at a height of 6 m and a distance of 40 m between the street lamps. These cooperative perception RSUs seem in complex urban city environments indispensable, in order to detect hidden objects, unequipped vehicles, bicycles, pedestrians, etc. From our point of view, the described V2V/V2X system and therefore mmWave will play an important role in Intelligent Transport Systems (ITS) in addition to 700 MHz and 5.9 GHz frequency bands. In systems beyond 5G, Unmanned Aerial Vehicles (UAV) may use a similar concept for automated flying at low altitude, as shown in Figure 23.

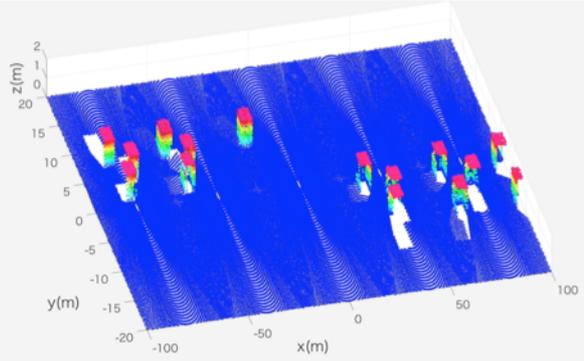

**Figure 25 Cooperative perception created by multiple RSUs on a road (This figure will be replaced with more realistic figure in the camera ready manuscript).**

A distinguishing feature of V2X versus other applications is the potential application of sensing on the RSU. Because the RSU is typically placed at a higher elevation, e.g. on a lamp pole, it has the advantage of supporting a birds-eye-view in a native fashion. This solves a key challenge in conventional systems where the sensing range of a vehicle may be obstructed due to the presence of a large truck or other obstruction. It also offers several other advantages. First, it becomes easier to monitor vehicles, bicycles, and pedestrians who are not equipped with V2X technology. Second, sensing information can be used to aid in establishing the mmWave communication link [71]. For example, an RSU mounted radar unit may be used to help track potential recipients of a mmWave communication exchange, thus reducing the beam alignment time [74]. Third, sensing at the RSU provides a source of sensor data for cities. This provides a new revenue stream and may also lead to further sharing of city-owned infrastructure in exchange for data access. The benefits could expand if additional sensors are deployed, including weather or pollution.

6.3 Spectrum Regulation in mmWave Frequency for Automated Driving

At the final part of this section, we will discuss the frequency spectra to be used for the mmWave V2V/V2X. Table 4 summarizes frequency candidates recently being selected for 5G and beyond in four different organizations, i.e. WRC-15 in ITU-R [12], CEPT in EU [17], FCC in US [14], and 5GMF in Japan [16]. Please note that the candidates in CEPT, FCC, and 5GMF were announced after the WRC-15 to be harmonized with ITU-R as much as possible. The frequency bands to be used for V2V/V2X should fulfill certain requirements, such as 1) have a bandwidth more than 1 GHz in order to exchange HD map information, 2) work internationally and regardless of country borders and regulatory bodies, namely use International Mobile Telecommunications (IMT) bands, 3) should be license-based to avoid unnecessary interference, and 4) should work standalone when Public Land Mobile Networks (PLMNs) are unavailable. Based on the above mentioned criteria, this paper nominates four candidates for mmWave V2V/V2X: (1) 31.8 – 33.4 GHz, (2) 40.5 – 42.5 GHz, (3) 47.0 – 50.2 GHz, and (4) 66.0 – 71.0 GHz. Although band (4) is recently regulated for unlicensed use in US, it is kept as a candidate because real applications in this band are still open to practical markets. The 28 GHz band is another candidate from the viewpoint of device availability, though the bandwidth is limited to 400 MHz per channel in US and there is no harmonization in the world. We have not selected the 71.0 – 76.0 GHz and 81.0 – 86.0 GHz bands to avoid interference with existing or future backhaul / fronthaul networks using these frequency bands.

**Table 4 Frequency candidates for 5G and beyond selected in the four different organizations.**

| WRC-15 | CEPT | FCC | 5GMF |
|---|---|---|---|
| 24.25 – 27.5 | 24.25 – 27.5 | | |
| | | | 24.75 – 31.0 |
| | | 27.5 – 28.35 | |
| 31.8 – 33.4 | 31.8 – 33.4 | 31.8 – 33.4 | |
| 37.0 – 40.5 | | 37.0 – 38.6 | 31.5 – 42.5 |
| | | 38.6 – 40.0 | |
| 40.5 – 42.5 | 40.5 – 43.5 | 40.5 – 42.5 | |
| 42.5 – 43.5 | | | |
| | | | 45.3 – 47.0 |
| 45.5 – 47.0 | | | |
| 47.0 – 47.2 | 45.5 – 48.9 | | |
| | | | 47.0 – 50.2 |
| 47.2 – 50.2 | | 47.2 – 50.2 | |
| 50.4 – 52.6 | | | 50.4 – 52.6 |
| | | 64.0 – 71.0 | |
| 66.0 – 76.0 | 66.0 – 71.0 | | 66.0 – 76.0 |
| | 71.0 – 76.0 | 71.0 – 76.0 | |
| 81.0 – 86.0 | 81.0 – 86.0 | 81.0 – 86.0 | 81.0 – 86.0 |

## 7. Concluding Remarks

Answers to the questions of the title regarding "where, when, and how mmWave is used in 5G and beyond" can be summarized as followed. Firstly, 28 GHz mmWave band will be used as backhaul network for moving hotspots, such as buses, to showcase the world first 5G entertainment system during the 2018 PyeongChang Winter Olympics in South Korea. Concurrently, the 28

GHz band will be used in the US. Secondly, 60 GHz mmWave band will be combined with MEC and used in 2018 for on-demand content download for smartphones and tables at Tokyo-Narita airport, in order to provide Omotenashi services for guests coming to the 2020 Tokyo Summer Olympic. The combination of mmWave and MEC is the only way to satisfy both extreme communications requirements: ultra-high speed and low latency. The third deployment phase around 2020 involves mmWave mesh networks, which will be deployed as an integrated and cost-effective access and backhaul system in dense urban scenarios. Additional traffic and energy management algorithms based on SDN technology will further decrease the cost of mmWave mesh network operations. After 2020, mmWave based V2V/V2X services will be deployed, in order to realize automated driving in complex urban environments. This is accomplished by cooperative perception and the exchange of HD dynamic map information between vehicles and RSUs, in order to enhance the visibility area. The automated driving use case can be considered as the most important application of mmWave and MEC, which requires both ultra-high speed and low latency. We believe that more mmWave and MEC native applications will emerge beyond 2020.

**Acknowledgments**

The research leading to these results received funding from the following research programs.
• European Commission H2020 programme under grant agreement N°723171 (5G-MiEdge project)
• The Ministry of Internal Affairs and Communications, Japan under grant agreement N°0159-0149, N°0159-0150, N°0159-0151 (5G-MiEdge project)
• The Ministry of Internal Affairs and Communications, Japan under grant agreement N°0155-0164 (MiEdge+ project)
• European Commission H2020 programme under grant agreement N°723247 (5G Champion project)
• Institute for Information & communication Technology Promotion (IITP) grant funded by the Korea government (MSIP) under grant agreement N°B0115-16-0001 (5G Champion project).

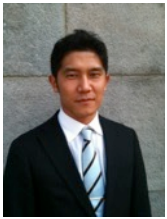
**Kei Sakaguchi** received the M.E. degree in Information Processing from Tokyo Institute Technology in 1998, and the Ph.D degree in Electrical & Electronics Engineering from Tokyo Institute Technology in 2006. Currently, he is working at Fraunhofer HHI in Germany as a Senior Scientist and at the same time he is an Associate Professor at Tokyo Institute of Technology in Japan. He received the Outstanding Paper Awards from SDR Forum and IEICE in 2004 and 2005 respectively, and three Best Paper Awards from IEICE communication society in 2012, 2013, and 2015. He also received the Tutorial Paper Award from IEICE communication society in 2006. He served as a TPC co-chair in the IEEE 5G Summit in 2016, a General co-chair in the IEEE WDN-5G in 2017, and a Industrial Workshop co-chair in the IEEE Globecom in 2017. His current research interests are in 5G cellular networks, millimeter-wave communications, and wireless energy transmission. He is a member of IEICE and IEEE.

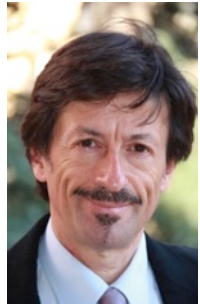
**Sergio Barbarossa** received his MS and Ph.D. EE degree from the Sapienza University of Rome, where he is now a Full Professor. He has held visiting positions at the Environmental Research Institute of Michigan ('88), Univ. of Virginia ('95, '97), and Univ. of Minnesota ('99). He is an IEEE Fellow, EURASIP Fellow, and he has been an IEEE Distinguished Lecturer. He received the IEEE Best Paper Awards from the IEEE Signal Processing Society for the years 2000 and 2014. He received the Technical Achievements Award from the European Association for Signal Processing (EURASIP). He has written a book entitled Multiantenna Wireless Communication Systems. He is currently a member of the editorial board of the *IEEE Transactions on Signal and Information Processing over Networks*. He has been the scientific coordinator of several EU projects on wireless sensor networks, small cell networks, and distributed mobile cloud computing. He is now the technical manager of the H2020 Europe/Japan project 5G-MiEdge. His current research interests are in the area of graph signal processing, distributed optimization, millimeter wave communications, mobile edge computing and machine learning.

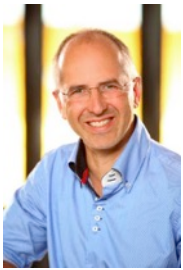
**Thomas Haustein** received the Dr.-Ing. (Ph.D.) degree in mobile communications from the University of Technology Berlin, Germany, in 2006. In 1997, he was with the Fraunhofer Institute for Telecommunications, Heinrich Hertz Institute (HHI), Berlin, where he worked on wireless infrared systems and radio communications with multiple antennas and OFDM. He focused on real-time algorithms for baseband processing and advanced multiuser resource allocation. From 2006 till 2008, he was with Nokia Siemens Networks, where he conducted research for 4G. Since 2009 he is heading the Wireless Communications Department at Fraunhofer HHI focussing on 5G and Industrial Wireless and related standardization. He was coordinating the Europe/Japan project MiWEBA and is currently coordinating the European team of the H2020 Europe/Japan project 5G-MiEdge.

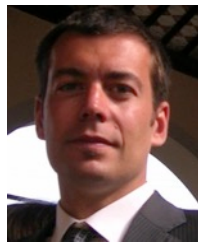
**Emilio Calvanese Strinati** obtained his Engineering Master degree in 2001 from the University of Rome 'La Sapienza' and his Ph.D in Engineering Science in. He then started working at Motorola Labs in Paris in 2002. Then in 2006 he joint CEA/LETI as a research engineer. From 2007, he becomes a PhD supervisor. From 2010 to 2012, Dr. Calvanese Strinati has been the co-chair of the wireless working group in GreenTouch Initiative which deals with design of future energy efficient communication networks. From 2011 to 2016 he was the Smart Devices & Telecommunications European collaborative strategic programs Director. Since December 2016 he is the Smart Devices & Telecommunications Scientific and Innovation Director. In December 2013 he has been elected as one of the five representative of academia and research center in the Net!Works 5G PPP ETP. He is currently one of the three moderators of the 5G future network expert group. Since 2016 he is the coordinator of the H2020 joint Europe and South Korea 5G CHAMPION project.

  E. Calvanese Strinati has published around 100 papers in international

conferences, journals and books chapters, given more than 50 international invited talks and keynotes and is the main inventor or co-inventor of more than 60 patents. He has organized more than 50 international conferences, workshops, panels and special sessions on green communications, heterogeneous networks and cloud computing hosted in international conferences as IEEE GLOBCOM, IEEE PIMRC, IEEE WCNC, IEEE VTC, EuCnC, IFIP, and European Wireless.

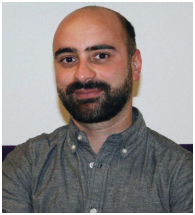

Antonio Clemente received the B.S. and M.S. degrees in telecommunication engineering and remote sensing systems from the University of Siena, Italy, in 2006 and 2009, respectively, and the Ph.D. degree in signal processing and telecommunications from the University of Rennes 1, France, in 2012. From 2008 to 2009, he realized his master's thesis project with the Technical University of Denmark, Lyngby, Denmark, where he was involved in spherical near-field antenna measurements. His Ph.D. degree has been realized at CEA, LETI, Grenoble, France. In 2012, he joined the Research and Development Laboratory, Satimo Industries, Villebon-sur-Yvette, France. Since 2013, he has been a Research Engineer at CEA, LETI. His current research interests include fixed-beam and electronically reconfigurable transmitarray antennas at microwave and millimeter-wave frequencies, antenna arrays, miniature integrated antennas, antenna fundamental limitations, near-field, and far-field antenna measurements. He serves as a reviewer for numerous IEEE and IET journals in the field of microwave, antennas, and propagation. Dr. Clemente was a recipient of the Young Scientist Award (First Prize) during the 15th International Symposium of Antenna Technology and Applied Electromagnetics held in 2012 and a co-recipient of the Best Paper Award at 19emes Journées Nationales Microondes held in 2015.

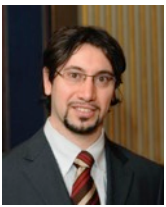

**Giuseppe Destino** received his Dr. Sc. degree at the University of Oulu in 2012 and, two M.Sc. (EE) degrees simultaneously from the Politecnico di Torino, Italy, and the University of Nice, France, in 2005. He received the Master of Research in Digital Signal Processing from the University of Nice and the Research Institute EURECOM, Sophia-Antipolis, France. Currently he is working as a Academy of Finland postdoctoral researcher as well as project manager of national and international projects at the Centre for Wireless Communications of the University of Oulu, Finland. His research interests include wireless communications with special emphasis on millimeter wave radio access technologies. He has co-authored three book chapters as well as several publications in the area of channel estimation, hybrid beamforming and positioning. His research work has been cited in more than 380 publications. He served as a member of the technical program committee of IEEE conferences.

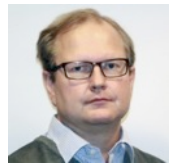

**Aarno Pärssinen** received the Doctor of Science degrees in electrical engineering from the Helsinki University of Technology, Finland, in 2000. In 1996, he was a Research Visitor at the University of California at Santa Barbara. From 2000 to 2011 he was with Nokia Research Center, Helsinki, Finland where he served as a member of Nokia CEO Technology Council from 2009 to 2011. From 2011 to 2013, he was at Renesas Mobile Corporation, Helsinki, Finland and then joined Broadcom, Helsinki, Finland as part of business acquisition until September 2014. Since September 2014 he has been with University of Oulu, Centre for Wireless Communications, Finland where he is currently a Professor. His research interests include wireless systems and transceiver architectures for wireless communications with special emphasis on the RF and analog integrated circuit and system design. He has authored and co-authored one book, one chapter of a book, more than 50 international journal and conference papers and holds several patents. He is also one of the original contributors to Bluetooth low energy extension, now called as BT LE. He served as a member of the technical program committee of Int. Solid-State Circuits Conference in 2007-2017, chairing the wireless subcommittee in 2014-2017.

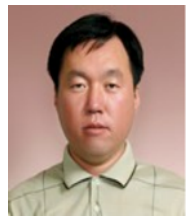

**Ilgyu Kim** received his BS and MS degrees in Electronic Engineering from University of Seoul, Rep. of Korea, in 1993 and 1995, and his PhD degree in Information Communications Engineering from KAIST 2009. Since 2000, he has been with ETRI, where he has been involved in the development of WCDMA, LTE and MHN systems. Since 2012, he has been the leader of the mobile wireless backhaul research section. His main research interests include millimeter wave communications and 5G mobile communications.

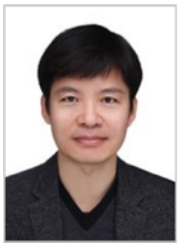

**Heesang Chung** received his BS in physics from the Korea Advanced Institute of Science and Technology, Daejeon, Rep. of Korea, in 1993, and the MS and PhD degrees from Chungnam National University, in 1995 and 1999, respectively. Since then, he has been with the ETRI where he is currently the Principal Researcher. His career at ETRI began with optical communications, and moved on to mobile and wireless communication systems in 2005. He was involved in research projects related to LTE and LTEAdvanced from 2006 to 2010. His recent research interests are in 5G with a special emphasis on high data-rate services for passengers on public transportation, such as the buses, subway and bullet trains.

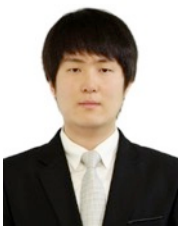

**Junhyeong Kim** received his B.S. degree in Dept. of Electronic Engineering from Tsinghua University, Beijing, China, in 2008, and M.S. degree in Dept. of Electrical Engineering from Korea Advanced Institute of Science and Technology (KAIST), Korea, in 2011. Since 2011, he has been with Electronics and Telecommunications Research Institute (ETRI), Korea. He is also currently pursuing his Ph.D. degree in School of Electrical Engineering at KAIST. His main research interests include millimeter-wave communications, MIMO, cooperative communications, and handover.

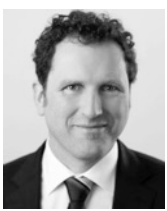

**Wilhelm Keusgen** received the Dipl.-Ing. (M.S.E.E.) and Dr.-Ing. (Ph.D.E.E.) degrees from the RWTH Aachen University, Aachen, Germany, in 1999 and 2005, respectively. From 1999 to 2004, he was with the Institute of High Frequency Technology, RWTH Aachen University. Since 2004 he is heading a research group for mm-waves and advanced transceiver technologies at the Fraunhofer Heinrich Hertz Institute, located in Berlin, Germany. His main research areas are millimeter wave communications for 5G, measurement and modeling of wireless propagation channels, multiple antenna systems, and compensation of transceiver impairments. Since 2007 he also has a lectureship at the Technical University Berlin.

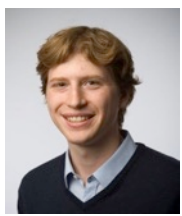

**Richard J. Weiler** received his diploma (Dipl.-Ing.) in electrical engineering and a diploma in business administration (Dipl.-Wirt. Ing.) from RWTH Aachen University in Germany. In 2016 he received the Dr.-Ing. (Ph.D.) degree from Technical University of Berlin for his work on the investigation of millimeter-wave outdoor access channels.
In 2010 he joined the Fraunhofer Heinrich-Hertz-Institute in Berlin, Germany, where he is currently working as project manager for the EU-Japan project MiEdge. His main research interest is millimeter-wave communication with a focus on propagation and the lower network layers as well as system level design of millimeter-wave communication systems.

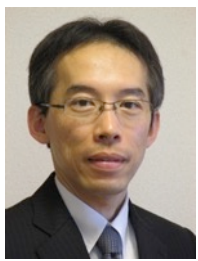

**Koji Takinami** received the B.S. and M.S. degrees in electrical engineering from Kyoto University, Kyoto, Japan, in 1995 and 1997, respectively, and the Ph.D. degree in physical electronics from Tokyo Institute of Technology, Tokyo, Japan, in 2013. In 1997, he joined Matsushita Electric Industrial (Panasonic) Co., Ltd., Osaka, Japan. Since then he has been engaged in the design of analog and RF circuits for wireless communications. From 2004 to 2006, he was a visiting scholar at the University of California, Los Angeles (UCLA), where he was involved in the architecture and circuit design of the high efficiency CMOS power amplifier. In 2006, he joined Panasonic Silicon Valley Lab, Cupertino, CA, where he worked on high efficiency transmitters and low phase-noise digital PLLs. In 2010, he relocated to Japan and currently leads millimeter wave system development. Dr. Takinami is a co-recipient of the best paper award at the 2012 Asia-Pacific Microwave Conference, the best invited paper award from IEICE Electronics Society in 2015 and the Electrical Science and Engineering Promotion Awards (the OHM Technology Award) in 2015. He was a member of the ISSCC Technical Program Committee from 2012 to 2014.

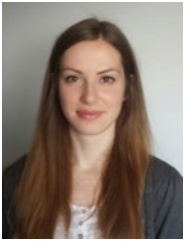

**Elena Ceci** received her BS and MS degree in electronics engineering from Sapienza University of Rome in 2014 and 2016, respectively. She is currently pursuing her Ph.D degree in Information and Communications Technologies at Sapienza University of Rome. Her research interests are in the area of mobile edge computing, millimeter-wave communications and machine learning.

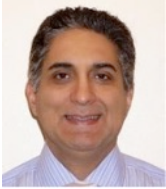

**Ali Sadri** is Sr. Director of mmWave Standards and Advanced TechnologiesAli Sadri is Sr. Director of mmWave Standards and Advanced Technologies at Intel Corporation. He has over 25 years of engineering, Scientific and academic background in Wireless Communications system. His Professional work started at IBM in conjunction with serving as a Visiting Professor at the Duke University. In 2002 Ali joined Intel Corporation where he initiated and lead the standardization of the next generation High Throughput WLAN at Intel that became the IEEE 802.11n standards. Later Dr. Sadri founded and served as the CEO of the Wireless Gigabit Alliance in 2008 that created the ground breaking WiGig 60 GHz technology and the 802.11ad standards. Later in 2013 WiGig Alliance merged with WiFi Alliance to advance and certify the WiGig programs within WiFi alliance framework. Currently Dr. Sadri is leading the mmWave advanced technology development that includes the next generation WiGig standards and mmWave technology for enabling 5G systems utilizing mmWave for Backhaul, Fronthaul and distribution systems. Ali holds over 100 Issued patents in wired and wireless communications systems. at Intel Corporation. He has over 25 years of engineering, Scientific and academic background in Wireless Communications system. His Professional work started at IBM in conjunction with serving as a Visiting Professor at the Duke University. In 2002 Ali joined Intel Corporation where he initiated and lead the standardization of the next generation High Throughput WLAN at Intel that became the IEEE 802.11n standards. Later Dr. Sadri founded and served as the CEO of the Wireless Gigabit Alliance in 2008 that created the ground breaking WiGig 60 GHz technology and the 802.11ad standards. Later in 2013 WiGig Alliance merged with WiFi Alliance to advance and certify the WiGig programs within WiFi alliance framework. Currently Dr. Sadri is leading the mmWave advanced technology development that includes the next generation WiGig standards and mmWave technology for enabling 5G systems utilizing mmWave for Backhaul, Fronthaul and distribution systems. Ali holds over 100 Issued patents in wired and wireless communications systems.

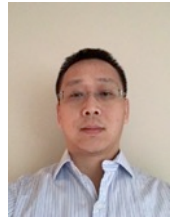

**Liang Xian** received the B.S. degree in electronics science from Nankai University, China in 1997, M.S. degree in telecom. engineering from Chongqing University of Posts and Telecommunications, China in 2000, and the Ph.D. degree in electrical engineering from Oregon State University, USA in 2006. From April 2013, he has been with Intel Corporation, where he is working on advanced mmWave technology.

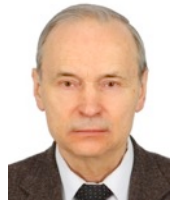

**Alexander Maltsev** received the Doctor of Science degree, in Radiophysics, from the University of Nizhny Novgorod (UNN), Russia, in 1990. After that he held the position of a Full Professor (1990 to 1994) and from 1994 till present he is holding the position of Head of the Statistical Radiophysics department in the UNN. In April 2001 Prof. Maltsev joined Intel Corporation and created Advance Development (AD) team in Intel Russia from his former PhD students. From April 2006 till present he is an Intel Principal Engineer managing AD team. During that time he contributed to development of Wi-Fi (IEEE802.11n, 11ad), WiMAX (IEEE802.16e, 16m), WiGig and LTE standards, participated in FP6 MEMBRANE (Multi-Element Multihop Backhaul Reconfigurable Antenna Network) project (2006-2009), headed the Russian Evaluation Group in IMT-Advanced (4G) evaluation process in ITU (2010-2011), participated as a technical expert in joint EU-Japan research FP7 project MiWEBA (Millimeter-Wave Evolution for Backhaul and Access) in 2013-2016. Prof. Maltsev is author of numerous papers in refereed journals and conferences proceedings and holds more than 100 US patents. His research interests include optimal and adaptive statistical signal processing, mmWave high-gain steerable antennas, MIMO-OFDM communication systems including 5G. Currently, he is working in IEEE802.11ay group on new PHY layer design and leads "Channel Models for IEEE802.11ay" document development.

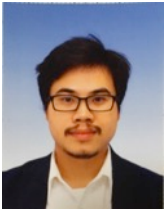

**Gia Khanh Tran** was born in Hanoi Vietnam, on February 18, 1982. He received the B.E., M.E., and D.E. degrees in electrical and electronic engineering from Tokyo Institute of Technology in 2006, 2008, and 2010. Currently, he is working as an Assistant Professor at the same university. He received the IEEE VTS Japan 2006 Young Researcher's Encouragement Award from the IEEE VTS Japan Chapter in 2006 and the IEICE Service Recognition Awards in 2013 and 2015. He also received the Best Paper Award in Software Radio from IEICE SR technical committee in 2009 and 2013, the Best Paper Award at SmartCom2015 and the Best Paper Awards from both IEICE and IEICE ComSoc in 2014. He served as a TPC co-chair in a series of IEEE WDN workshops including WDN-5G in ICC2017. His research interests are signal processing, MIMO mesh networks, coordinated heterogeneous cellular networks, mm-wave communication, localization. He is a member of IEEE and currently the assistant of the technical committee on Smart Radio (TCSR) of IEICE ComSoc.

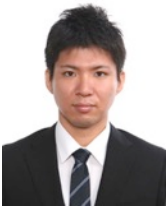

**Hiroaki Ogawa** was born in Tokyo, Japan, on December 5, 1993. He received the B.E. degree in clectrical and electronic engineering from Tokyo Institute of Technology, Japan, in 2016. From 2016, he is a Master's course student in the department of Electrical and Electronic Engineering, Tokyo Institute of Technology. His current research interest is mmWave meshed backhaul networks in mmWave overlay heterogeneous networks. He is a student member of IEICE.

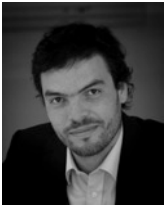

**Kim Mahler** received the M.Sc. degree with honors and the Dr.-Ing. degree from the Electrical Engineering and Computer Science Department, the Technical University of Berlin, Germany, in 2010 and 2016 respectively, and the M.A. degree from the Berlin University of Arts/University of St. Gallen in 2014. Since 2009, he is with the Wireless Communications and Networks department at Fraunhofer Heinrich Hertz Institute. His research interests involve user-centric 5G developments and the application of millimeter wave technology to vehicular and UAV communications scenarios.

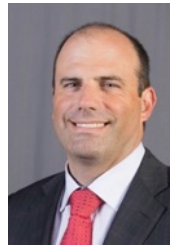

**Robert W. Heath Jr.** received the Ph.D. from Stanford University. Since January 2002, he has been with the Department of Electrical and Computer Engineering at The University of Texas at Austin where he is a Cullen Trust for Higher Education Endowed Professor, and is a Member of the Wireless Networking and Communications Group. He is also President and CEO of MIMO Wireless Inc. He authored "Introduction to Wireless Digital Communication" (Prentice Hall, 2017), co-authored "Millimeter Wave Wireless Communications (Prentice Hall, 2014), and authored "Digital Wireless Communication: Physical Layer Exploration Lab Using the NI USRP? (National Technology and Science Press, 2012). He has been a co-author of fifteen award winning conference and journal papers including recently the 2016 IEEE Communications Society Fred W. Ellersick Prize, and the 2016 IEEE Communications and Information Theory Societies Joint Paper Award. He was a distinguished lecturer in the IEEE Signal Processing Society and is an ISI Highly Cited Researcher. He is also an elected member of the Board of Governors for the IEEE Signal Processing Society, a licensed Amateur Radio Operator, a Private Pilot, and a registered Professional Engineer in Texas. He is a Fellow of the IEEE, and a member of EURASIP and IEICE.